# Spectral tunneling of lattice nonlocal solitons

Yaroslav V. Kartashov,[1] Victor A. Vysloukh,[2] and Lluis Torner[1]

[1]*ICFO-Institut de Ciencies Fotoniques, and Universitat Politecnica de Catalunya, Mediterranean Technology Park, 08860 Castelldefels (Barcelona), Spain*

[2]*Departamento de Fisica y Matematicas, Universidad de las Americas - Puebla, Santa Catarina Martir, 72820, Puebla, Mexico*

We address spectral tunneling of walking spatial solitons in photorefractive media with nonlocal diffusion of the nonlinear response and an imprinted shallow optical lattice. In contrast to materials with local nonlinearities, where solitons travelling across the lattice close to the Bragg angle suffer large radiative losses, in photorefractive media with diffusion nonlinearity resulting in self-bending solitons survive when their propagation angle approaches and even exceeds the Bragg angle. In the spatial frequency domain this effect can be considered as tunneling through the band of spatial frequencies centered around the Bragg frequency where the spatial group velocity dispersion is positive.

The phenomenon of soliton spectral tunneling, that was introduced for the first time for temporal solitons in optical fibers, manifests itself as a sharp transition of the soliton spectrum between two frequency domains where the group velocity dispersion is anomalous, separated by a domain with normal group velocity dispersion - in other words, a spectral gap where bright solitons cannot exist in an homogeneous focusing Kerr nonlinear medium. Such effect was predicted theoretically in the early 90s [1] and it has been recently demonstrated experimentally in specially designed dispersion-controlled holey fibers [2]. It was uncovered that a driving force of soliton spectral tunneling is the Raman self-frequency shift, which allows solitons to traverse the region where the group velocity dispersion is normal [3,4]. On the other hand, engineered materials may show unusual diffraction properties. Among them are photonic crystals and lattices that afford a wealth of new possibilities for spatial soliton control (see [5,6] for recent reviews). The propagation of light in such structures has been under intense investigation over the last decade and effects as reduction, cancellation, and even reversal of diffraction have been observed [7]. Important links be-tween band-gap structures in waveguide arrays [8] or photoinduced lattices [9,10] and domains of existence, symmetry, and stability of solitons have been established.

The possibility to engineer diffraction in periodic media suggests that an analog of temporal soliton spectral tunneling can be generated in the spatial domain. In this paper we introduce such an effect. We show that while spectral tunneling in the time domain is mediated by Raman self-frequency shift, spectral tunneling in the spatial domain may be driven by self-bending, a characteristic effects of photorefractive materials exhibiting quasi-local drift and nonlocal diffusion nonlinearity. Self-bending affects considerably the formation and interaction of spatial solitons in uniform photorefractive media [11-15], and it also allows soliton mobility control in the presence of optical lattices [16,17]. The key aspect of the approach that we put forward here is that self-bending results in a progressive shift of the central spatial soliton frequency (depending on the direction of biasing electric field and sign of electric charge of free carriers this shift can by negative or positive). When the soliton instantaneous propagation angle becomes such that diffraction is anomalous for the soliton central frequency (which is incompatible with existence of bright solitons in focusing media), the soliton still survives due to self-bending and tunnels a large part of its energy into the spectral region where diffraction becomes normal again. Note that this is in contrast to local media where solitons at such propagation angles would decay.

We consider the propagation of laser beam along the $\xi$ axis in a slab photorefractive waveguide with both local drift and nonlocal diffusion components of the nonlinear response and an imprinted periodic modulation of the refractive index along the transverse $\eta$ axis. Light propagation is described by the nonlinear Schrödinger equation for the dimensionless field amplitude $q$:

$$i\frac{\partial q}{\partial \xi} = -\frac{1}{2}\frac{\partial^2 q}{\partial \eta^2} - q|q|^2 + \mu q \frac{\partial}{\partial \eta}|q|^2 - pR(\eta)q. \qquad (1)$$

Here the longitudinal $\xi$ and transverse $\eta$ coordinates are scaled to the diffraction length and the characteristic beam width, respectively. The parameter $\mu$ describes the magnitude of the nonlocal diffusion component of the nonlinear response; $p$ is the guiding parameter that is proportional to refractive index modulation depth; the function $R(\eta) = \cos(\Omega \eta)$ describes the transverse profile of the refractive index; $\Omega$ is the lattice frequency (see [16] for details of normalizations). We consider shallow lattices with $p \leq 1$ and experimentally realistic values of the bending parameter $|\mu| \leq 0.3$. We also set $\Omega = 4$. Notice that in photore-fractive media the nonlinear contribution to the refractive index arises via the linear electro-optic effect and due to diffusion of free carriers, and that it is proportional to the internal space-charge field $\delta n \sim E_{\text{sc}} = [E_0 I_{\text{bg}} + (k_{\text{b}} T / e) \partial I / \partial x](I + I_{\text{bg}} + I_1)^{-1}$, where $I$ is the light intensity, $I_{\text{bg}}$ is the intensity of background illumination, $I_1$ describes intensity distribution in the beam that induces the lattice, $T$ is the temperature, $e$ is the charge of free carriers, $k_{\text{b}}$ is the Boltzman constant, and $E_0$ is the biasing field. Under the assumption $I + I_1 \ll I_{\text{bg}}$ the propagation of extraordinarily polarized light beams in such material can be described by Eq. (1) [12,14,16].

To understand the physics behind soliton spectral tunneling, it is instructive to consider the band-gap structure of the simplest linear harmonic lattice that may be used to characterize the light evolution in periodic media. Applying the substitution $q(\eta,\xi) = w(\eta) \exp[i(K\eta - b\xi)]$, where $b$ is the propagation constant, $K$ is the spatial frequency (Bloch momentum), and $w(\eta) = w(\eta + 2\pi / \Omega)$ is a periodic function, one can calculate from the linear version of Eq. (1) (at $|q|^2 \to 0$) the dispersion relations $b(K)$ for different bands where periodic Bloch-wave solutions exist.

Figure 1(a) shows dispersion relations $b_{1,2}(K)$ for the first two bands (the subscript indicates the number of the band). Notice a gap with a width $\sim 8^{1/2} p$ separating $b_1(K)$ and $b_2(K)$ curves in the vicinity of the Bragg frequency $K = \Omega / 2$. Using these dependencies one can show that a spatially broad (and spectrally narrow) beam featuring an internal structure similar to that of Bloch wave with momentum $K$ will propagate inside the lattice with a spatial group velocity $V = -\partial b / \partial K$, while the spatial group velocity dispersion (or diffraction) will be given by $D = \partial^2 b / \partial K^2$ [9]. The dependencies $D(K)$ for the first band are shown in Fig. 1(b) for different $p$ values. Notice that around $K = \Omega / 2$ a spectral barrier where $D(K) > 0$ (in this spectral region diffraction is anomalous, hence solitons can not exist in focusing media) appears that separates domains with $D(K) < 0$ (where diffraction is normal, hence solitons may exist) and that counteracts the transmission between spectral regions with low and high spatial frequencies. This barrier is mathematically analogous to the one introduced in the case of spectral tunneling of temporal solitons [1,4].

In the spatial frequency band where $D > 0$, the joint action of anomalous diffraction and focusing nonlinearity enhances the beam spreading. Importantly, the width of this barrier grows, while its height diminishes with increase of $p$ almost linearly, so that the quantity $\int D(K) dK$ remains nearly constant. However, since transmission through the barrier decreases more rapidly with increase of barrier width than with increase of its height, at a fixed area $\int D(K) dK$ the transmission of soliton through a wide and low barrier ($p=1$) is expected to be more difficult than through a narrow and high barrier $(p=0.25)$. While in local nonlinear media such spectral barrier prevents transmission between regions with high and low spatial frequencies, due to self-bending that causes continuous shift of soliton spectrum upon propagation, the soliton can traverse the barrier without being destroyed. This is the central result of this paper. It should be also stressed that while in the time domain qualitative and quantitative features of spectral tunneling are totally determined by fixed parameters of the Raman scattering line [1,2], in the spatial domain the spectral tunneling can be flexibly controlled due to possibility of engineering of diffraction by such well developed techniques as modulation of lattice amplitude or frequency.

To illustrate the space-domain spectral tunneling we solved Eq. (1) numerically with the initial conditions $q(\eta,0)=\chi\,\mathrm{sech}(\chi\eta)\exp(i\alpha\eta)$ that correspond to an exact soliton solution of the local nonlinear Schrödinger equation [i.e., Eq. (1) with $p,\mu=0$]. Such soliton is characterized by the form-factor $\chi$, while the parameter $\alpha$ describes the tilt of the phase front and simultaneously determines the central frequency in the spatial spectrum at $\xi=0$. Figure 2 illustrates the typical dynamics of spectral tunneling. The evolution of the intensity distribution $|q(\eta,\xi)|^2$ with distance for the case $\mu>0$ is shown in Fig. 2(a). A soliton with form-factor $\chi=1.2$ and central frequency $\alpha=1<\Omega/2$ was used as the initial condition. Such soliton is rather narrow, so that its width is comparable with the half lattice period, thus its spectral width $\delta K\approx 1.53$ exceeds substantially the spectral barrier width given by $\delta K\approx 0.2$ at $p=0.25$. Due to self-bending caused by diffusion nonlinearity the trajectory of the soliton center in the spatial domain is almost parabolic with its instantaneous propagation angle with respect to $\xi$ axis gradually increasing. Despite the pronounced radiative losses at $\xi<12$, the soliton does not decay and maintains its shape even when its instantaneous propagation angle becomes equal to the Bragg angle $\Omega/2$ and exceeds it. Thus, self-bending enables soliton passage through the spectral barrier. Figure 2(b) shows the corresponding transformation of the spectral intensity $S(K,\xi)=|q(K,\xi)|^2$. Notice that initially the soliton spectrum is located well below the Bragg frequency. Self-bending results in monotonic up-shifting of the central frequency with $\xi$, so that the high-frequency wing of the soliton spectrum starts overlapping with the frequency band (spectral barrier) where the spatial group velocity dispersion is positive. Then Bragg scattering comes into play. Narrow bright stripe at $K\simeq-\Omega/2$ whose intensity initially grows rapidly with $\xi$ and then saturates correspond to backward Bragg scattering, while less pronounced stripe that appears at $K\simeq+\Omega/2$ is linked with forward Bragg scattering. Despite rather strong spectrally-selective radiative losses the soliton spectrum remains bell-shaped after the passage through the spectral barrier with positive $D(K)$ centered at $K=\Omega/2$.

Figures 2(c) and 2(d) illustrate a similar phenomenon but when self-bending for the opposite sign of $\mu$ causes a down-shift of the soliton central frequency from the value $\alpha=3>\Omega/2$ that initially exceeds Bragg frequency. The radiative losses in this case are larger. The rate of shift of the spatial spectrum decreases considerably after tunneling. The radiative losses after tunneling also become substantially weaker (see also [18,19] for discussion of radiative losses for moving lattice solitons in local media). Notice also that if the launching angle $\alpha$ is far from the Bragg one, the propagation distance up to the point where spectral tunneling appears may increase considerably. This, in turn, results in more pronounced radiative losses and may even lead to suppression of tunneling.

Figure 3 illustrates the salient features of the phenomenon of spectral tunneling for the case of $\mu>0$ (up-shifting of spectrum) and $\alpha=1.5$. To quantify the transmission we introduce the spectral energy transferred through the barrier at a fixed propagation distance $\xi=20$, as $U_{\rm tr}(\xi)=\int_{\Omega/2}^{\infty}|q(K,\xi)|^2 dK$, while to quantify Bragg scattering upon transmission we define the reflected energy as $U_{\rm ref}(\xi)=\int_{-\infty}^{0}|q(K,\xi)|^2 dK$. Figure 3(a) shows the dependence of the reflected energy $U_{\rm ref}$ on the self-bending parameter $\mu$. This dependence is nonmonotonic because the propagation distance is fixed and for too small $\mu$ values the self-bending is not sufficient to shift soliton spectrum close to the region $K\sim\Omega/2$ where strong scattering occurs. For sufficiently large $\mu$, the growth of the self-bending parameter results in a decrease of $U_{\rm ref}$, i.e., faster tunneling is accompanied by smaller radiative losses. Figure 3(b) illustrates the dependence of the transmitted energy on $\mu$. The transmitted energy monotonically increases with $\mu$ and then saturates, indicating a reduction of backward Bragg scattering with $\mu$. In shallower lattices $U_{\rm tr}$ saturates more rapidly, mainly because of the reduced Bragg backward scattering [compare curves for $p=0.25$ and $p=0.35$ in Fig. 3(a)]. Figure 3(c) depicts the nonmonotonic dependence of reflected energy on soliton form-factor $\chi$. Because the rate of self-bending strongly depends on the form-factor, at relatively small $\chi$ values bending is insufficient to shift the soliton spectrum close to the spectral region causing large scattering. Thus, in this regime $U_{\rm ref}$ decreases with decreasing $\chi$. At large $\chi$ values the rate of bending grows and thus the soliton spectrum becomes broad in comparison with the width of the spectral barrier, resulting in a reduction of the backscattered energy. With increasing the lattice depth $p$, the form-factor corresponding to the maximum reflected energy increases. The transmitted energy increases with $\chi$ monotonically mainly because the input energy $U\sim\chi$ [Fig. 3(d)].

An interesting geometry where the phenomenon described here may have important implications is the so called Bragg-type soliton mirror, which was previously studied for the case of local cubic nonlinearity [20]. Soliton reflection and refraction at the interface formed by an uniform nonlinear medium (occupying the domain $\eta < 0$) and a periodic material (occupying the domain $\eta \geq 0$) offer rich opportunities for controlling the reflection efficiency and beam steering, because such reflection is strongly sensitive to variation in depth and frequency of the periodic refractive index modulation, as well as to the angle of incidence. In experiments with photorefractive materials the nonlocal diffusion component of nonlinear response usually appears as a perturbation ($|\mu| \leq 0.3$), but our simulations indicate that even for such small $\mu$ values the influence of the asymmetric diffusion nonlinearity on soliton reflection is considerable. In particular, if the incidence angle is close to Bragg one and self-bending effect causes deflection of soliton (that is launched inside uniform medium) into the depth of the periodic medium, the transmission is considerably enhanced and self-bending of transmitted beam inside the lattice can be observed. In contrast, for opposite sign of $\mu$, when self-bending occurs toward uniform medium the reflection becomes much stronger (notice that for $\mu = 0$ it is not 100% even at Bragg angle [20]).

Summarizing, we introduced the spatial analog of temporal soliton spectral tunneling in photorefractive media with an imprinted optical lattice. We showed that due to self-bending caused by diffusion in suitable photorefractive nonlinear media, spatial solitons moving across the lattice sites may tunnel the spectral region where diffraction is anomalous, hence regions where they would decay in local media with focusing nonlinearity.

# Figure captions

Figure 1 (color online). (a) The dependencies of propagation constant $b$ of Bloch wave on $K$ for first two bands in Floquet-Bloch spectrum. (b) The spatial group velocity dispersion $d^2b/dK^2$ versus $K$ for the first band. Dashed line in (b) corresponds to $d^2b/dK^2=0$. In all cases $\Omega=4$. All quantities are plotted in arbitrary dimensionless units.

Figure 2 (color online). Propagation dynamics (a) and evolution of spectrum (b) for $\alpha=1$, $\mu=0.25$, $\chi=1.2$. Propagation dynamics (c) and evolution of spectrum (d) for $\alpha=3$, $\mu=-0.25$, $\chi=1.2$. In all cases $\Omega=4$, $p=0.25$. All quantities are plotted in arbitrary dimensionless units.

Figure 3 (color online). Reflected (a) and transmitted (b) energy versus $\mu$ at $\alpha=1.5$, $\chi=1.2$. Reflected (c) and transmitted (d) energy versus $\chi$ at $\alpha=1.5$, $\mu=0.2$. In all cases $\Omega=4$. All quantities are plotted in arbitrary dimensionless units.

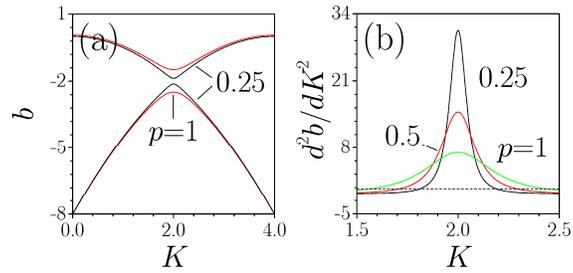

Figure 1 (color online). (a) The dependencies of propagation constant $b$ of Bloch wave on $K$ for first two bands in Floquet-Bloch spectrum. (b) The spatial group velocity dispersion $d^2b/dK^2$ versus $K$ for the first band. Dashed line in (b) corresponds to $d^2b/dK^2 = 0$. In all cases $\Omega = 4$. All quantities are plotted in arbitrary dimensionless units.

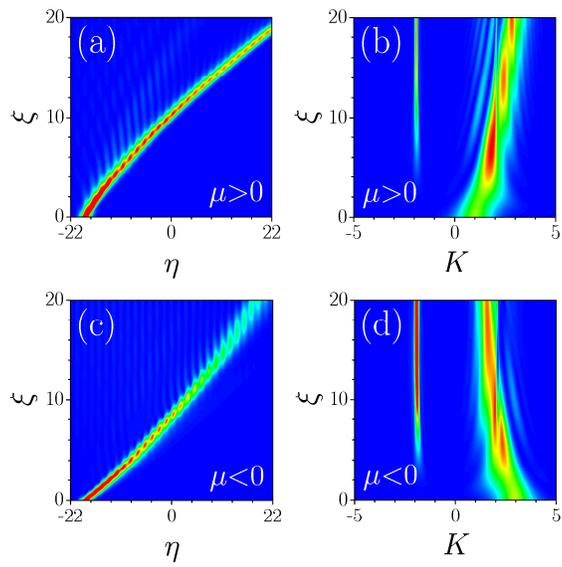

Figure 2 (color online). Propagation dynamics (a) and evolution of spectrum (b) for $\alpha=1$, $\mu=0.25$, $\chi=1.2$. Propagation dynamics (c) and evolution of spectrum (d) for $\alpha=3$, $\mu=-0.25$, $\chi=1.2$. In all cases $\Omega=4$, $p=0.25$. All quantities are plotted in arbitrary dimensionless units.

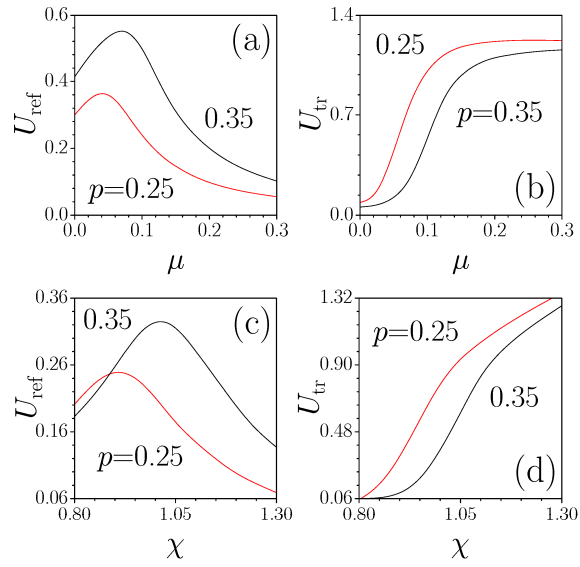

Figure 3 (color online). Reflected (a) and transmitted (b) energy versus $\mu$ at $\alpha = 1.5$, $\chi = 1.2$. Reflected (c) and transmitted (d) energy versus $\chi$ at $\alpha = 1.5$, $\mu = 0.2$. In all cases $\Omega = 4$. All quantities are plotted in arbitrary dimensionless units.